\documentclass{article}
\usepackage{dcase2023,amsmath,graphicx,url,times,booktabs, tabularx}

\usepackage{amsfonts} 
\usepackage{amsmath,amsfonts,amssymb}
\usepackage{fixltx2e}
\usepackage{multirow}
\usepackage{hhline}
\usepackage{graphicx}
\usepackage{subcaption}
\usepackage{mwe}

\title{Weakly-supervised Automated Audio Captioning via text only training}

\twoauthors
  {Theodoros Kouzelis}
    {Institute for Language and Speech Processing\\
        Athena Research Center\\ 
       15125, Marousi, Greece \\
     theodoros.kouzelis@athenarc.gr}
  {Vassilis Katsouros}
    {Institute for Language and Speech Processing\\
        Athena Research Center\\ 
       15125, Marousi, Greece \\
     vsk@athenarc.gr}

\begin{document}

\ninept
\maketitle

\begin{sloppy}

\begin{abstract}
In recent years, datasets of paired
audio and captions have enabled remarkable success in
automatically generating descriptions for audio clips, namely Automated Audio Captioning (AAC). However, it is labor-intensive and time-consuming
to collect a sufficient number of paired audio and captions. 
Motivated by the recent advances in Contrastive Language-Audio Pretraining (CLAP), we propose a weakly-supervised approach to train an AAC model assuming only text data and a pre-trained CLAP model, alleviating the need for paired target data. 
Our approach leverages the similarity between audio and text embeddings in CLAP. During training, we learn to reconstruct the text from the CLAP text embedding, and during inference, we decode using the audio embeddings.
To mitigate the modality gap between the audio and text embeddings we employ strategies to bridge the gap during training and inference stages. We evaluate our
proposed method on Clotho and AudioCaps datasets demonstrating its ability to achieve a relative performance of up to ~$83\%$ compared to fully supervised approaches trained with paired target data. \footnote{This work was conducted in the  framework of the PREMIERE project (No. 101061303) that is funded by the European Union.}  Our code is available at: \url{https://github.com/zelaki/wsac}
\end{abstract}


\section{Introduction}
\label{sec:intro}

Audio-Language tasks have recently gained the attention of the audio community with the introduction of Automated Audio Captioning and Language-Based Audio Retrieval in the DCASE Challenge and the release of publicly available Audio-Language datasets such as Clotho \cite{clotho} and AudioCaps \cite{audiocaps}.
The intrinsic relationship between Audio and Language presents an opportunity for the development of models that can effectively establish a shared semantic space for the two modalities. Such an approach has recently achieved great success
with models like COALA \cite{coala}, AudioClip \cite{audioclip}, and CLAP \cite{clap1,clap2,clap3}. These models use parallel audio-text data to train a joint representation,
where the embeddings of audio-text pairs are similar. Such models achieve high accuracy in a zero-shot setting in a variety of tasks including Sound Event Classification, Music tasks,
and Speech-related tasks \cite{clap1}.

Automated Audio Captioning (AAC) is a multimodal task that aims
to generate textual descriptions for a given audio clip. In order
to generate meaningful descriptions, a method needs to capture
the sound events present in an audio clip and generate a description in natural language.
Training audio captioning models requires large datasets of audio-caption pairs, and these are challenging to collect.
While great effort has been done, the data scarcity issue of audio captioning still withholds.
The common datasets in AAC,
AudioCaps and Clotho, contain together ~50k captions for training, whereas ~400k captions are provided in COCO caption \cite{coco} for image captioning. Kim et al. \cite{prefix} observe that due to the limited data, prior arts design decoders with shallow layers that fail to learn generalized language expressivity and are fitted to the small-scaled target dataset. Due to this issue, their performance radically decreases when tested on out-of-domain data.
Motivated by these limitations we present an
approach to AAC that only requires a pre-trained CLAP model and
unpaired captions from a target domain.
This alleviates the need for paired audio-text data, and
also allows for simple and efficient domain adaptation.

Our approach is inspired by recent advances in zero-shot image captioning \cite{noise, decap}, that leverage the aligned multi-modal latent space provided by CLIP \cite{clip} obviating the need for image data during training and by the recent success of Contrastive Language-Audio models such as CLAP \cite{clap1} in many downstream tasks. 
We train a lightweight decoder model to reconstruct texts
from their respective CLAP embeddings, and at inference use this decoder to decode the audio embeddings. Our findings align with prior studies in image captioning suggesting that such an approach is suboptimal due to the
presence of a phenomenon known as \textit{modality gap} \cite{gap}.

The \textit{modality gap} suggests that embeddings from different data modalities are located in two completely separate regions of the embedding
space of multi-modal contrastive models \cite{gap}. To mitigate this issue we employ strategies that have been shown to effectively condense the gap in CLIP embeddings \cite{noise, decap} and show that they can be effectively utilized for CLAP models. 
These strategies can be divided into two categories, strategies that condense the gap during \textit{training} and during \textit{inference}.

Experiments on Clotho and AudioCaps
datasets show that our weakly-supervised approach can achieve comparable performance to prior fully supervised arts, without requiring any target audio data during training. Our contributions can be summarized as follows: (1) We propose \textbf{WSAC:} \textbf{W}eakly-\textbf{S}upervised \textbf{A}udio \textbf{C}aptioning  an AAC approach that requires no auditory in-domain data for training, (2) we demonstrate that the \textit{modality gap} phenomenon is present in CLAP models, and (3) employ methods that effectively mitigate it.

\begin{figure*}[t]
  \centering
  \caption{Overview of our proposed approach. \textbf{Left:} An illustration of the CLAP training paradigm.
The encoders are trained to map semantically similar audio-caption pairs to similar embeddings in a joint representation space. \textbf{Middle:} Our proposed weakly supervised training. A frozen CLAP text encoder embeds a caption and a decoder learns to reconstruct the caption from its embedding. \textbf{Right:} At inference, we decode the audio embedding extracted from a frozen CLAP audio encoder, using the trained decoder.}
  \centerline{\includegraphics[width=0.85\textwidth, height=6.5cm]{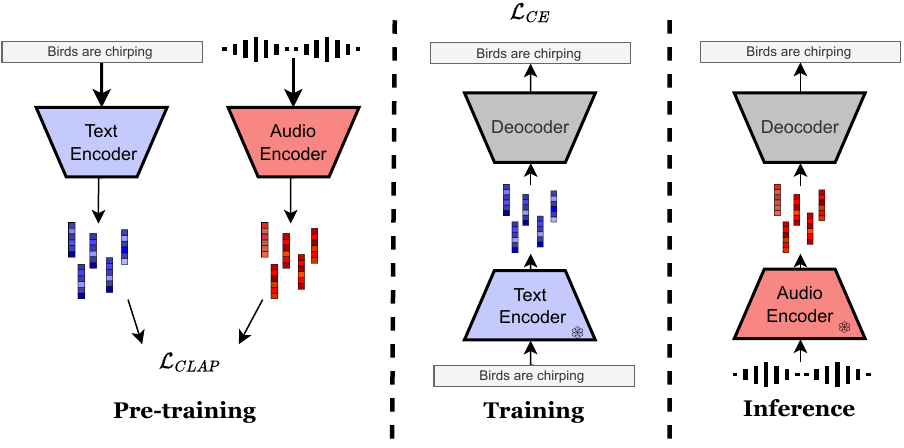}}
  \label{fig:main}
  \vspace{-10pt}
\end{figure*}

\section{Text-only training}
\label{text_training}
Our goal is to learn a model
that produces a caption for a given audio clip. Unlike fully supervised approaches, 
 during
training we only assume that we have access to a set of target domain captions $\mathcal{C}$.
We further assume a pre-trained CLAP model
with an audio encoder $\mathcal{A}_{clap}$ and a text encoder $\mathcal{T}_{clap}$ trained to project semantically similar audio-text pairs into similar embeddings in a shared embedding space as presented in Fig. \ref{fig:main} (Left).
Given an audio clip $x_a$ and text $x_t$ let $\mathbf{z_a}=\mathcal{A}_{clap}(x_a) \in \mathbb{R}^d$ and $\mathbf{z_t}=\mathcal{T}_{clap}(x_t) \in \mathbb{R}^d$ be their embeddings.

First we extract text embeddings $\mathbf{z}_t$ for all $x_t \in \mathcal{C}$, keeping  $\mathcal{T}_{clap}$ frozen.
During training, our goal is to learn a network that inverts the CLAP text encoder $\mathcal{T}_{clap}$.
We use a textual decoder $D$ consisting
of a mapping network $f$ and an
auto-regressive language model, to reconstruct the original text $x_t$ from the CLAP text embedding $\mathbf{z_t}$.
Following recent work \cite{prefix}, we train our decoder using the prefix language modeling paradigm.
Specifically, after passing the text embedding through the mapping network $f$ we regard $\mathbf{p} = f(\mathbf{z_t})$ as a prefix to the caption. Given a text $t = \{w_1, w_2, ..., w_{T}\}$, our objective is to minimize the autoregressive cross-entropy loss:
\begin{equation}
        \label{loss}
        \mathcal{L} =  - \sum_{i=1}^T \log D(w_i \vert w_{<i}, \mathbf{p})
\end{equation}

Since the CLAP text embedding is optimized
to be similar to the CLAP audio embedding, we can directly infer the text decoder
using the audio embeddings $\mathbf{z}_a$ without any pairwise training on the target dataset.
The training and inference stages are presented in Fig. 1 (middle) and (right) respectively.

\section{Stradegies to bridge the modality gap}
Directly employing the audio embeddings to infer $D$ is not optimal due to the presence of the modality gap. Fig. 2 is a visualization of generated embeddings from the pre-trained CLAP model from the Clotho training set. Paired inputs are fed into the pre-trained model and
the embeddings are visualized in 2D using T-SNE \cite{tsne}. This visualization clearly demonstrates the presence of the modality gap phenomenon, as a noticeable gap separates the paired audio and text embeddings. 
To address this issue, we utilize strategies that have demonstrated success in bridging the modality gap in CLIP embedding space \cite{noise, decap, gap}. We show that these strategies can be adopted for CLAP and show their effectiveness in mitigating the modality gap. These approaches can be divided into two categories: Bridging the gap either during the training phase or during the inference phase.

\subsection{Training strategies}
\label{training str}
Attempting to reduce the modality gap during training we adopt the following strategies: (a) Noise injection \cite{noise}, and Embedding Shift \cite{gap}. These strategies aim to narrow the disparity between the modality used to train the decoder, which is text, and the target modality, which is audio.

\subsubsection{Noise injection}
\label{noise}
In \cite{noise}, the authors show  that injecting the text embedding with Gaussian noise during training 
has the effect of creating a region in the embedding space that
will map to the same caption. This method assumes that the corresponding audio embedding is more likely to be inside this
region. Following \cite{noise}, we add zero-mean Gaussian noise of  standard deviation $\sigma$ to
the text embedding before feeding it to the decoder. 
We set $\sigma$ to the mean $L_{inf}$ norm of
embedding differences between five captions that
correspond to the same audio. Since we assume no access to target audio data we estimate $\sigma$ using 50 audio-caption pairs from the WavCaps dataset \cite{clap3}. Thus the prefix in Eq. \ref{loss} becomes $\mathbf{p} = f(\mathbf{z_t} + \mathbf{n})$, where $\mathbf{n} \in \mathbb{R}^d$
is a random standard Gaussian
noise with standard deviation $\sigma$.

\begin{figure}[t]
  \centering
  \centerline{\includegraphics[width=0.75\columnwidth]{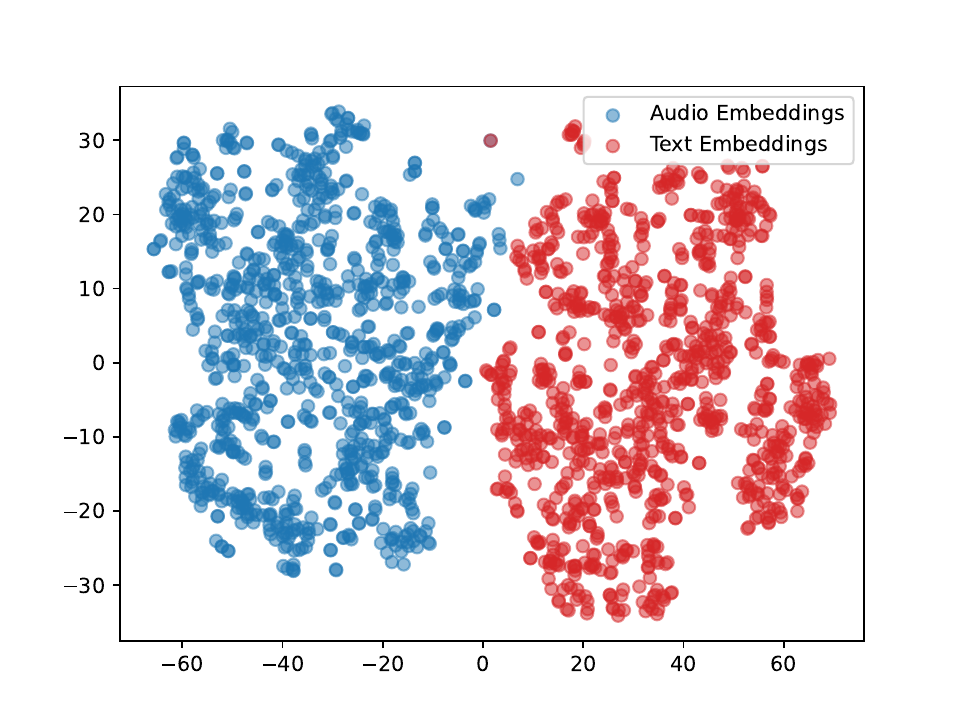}}
  \caption{Visualization of audio and text embedding pairs randomly sampled from the Clotho training set. The modality gap phenomenon is present as the audio and text modalities are embedded in two completely separate regions.}
  \label{fig:results}
  \vspace{-10pt}
\end{figure}

\subsubsection{Embedding shift}
\label{shift}
Building upon the findings of \cite{gap}, who investigated the impact of shifting embeddings in various multi-modal contrastive learning models on downstream tasks, we propose a method to align the text embeddings with the audio embeddings during training.
 First, we define the modality 
gap following \cite{gap}, as the difference between the center of audio embeddings and text embeddings:

\begin{equation}
    \mathbf{\Delta_{gap}} = \frac{1}{n} \sum_{i=1}^n \mathbf{z_{a_i}}  - \frac{1}{n} \sum_{i=1}^n  \mathbf{z_{t_i}} 
\end{equation}
Then, we 
shift every text embedding toward closing the modality gap, and thus the prefix in Eq. \ref{loss} becomes $ \mathbf{p}= f(\mathbf{z_t} + \mathbf{\Delta_{gap}})$.

\subsection{Inference strategies}
\label{inf str}
At inference, we adopt two training-free
strategies proposed in \cite{decap}, and map an audio embedding extracted from the CLAP audio encoder $\mathcal{A}_{clap}$ into  the text embedding space. For both strategies, we will assume a decoder $D$ trained on some target data as described in Section 2 and a set of text embeddings obtained from the target training set that we will refer to as \textit{Memory}, $\mathcal{M} = \{ \mathbf{z_t^1}, \mathbf{z_t^2}, ... \mathbf{z_t^N} \}$, where $N$ is the size of the training set.

\subsubsection{Nearest-neighbor decoding}
\label{nnd}
A straightforward strategy that can be adopted at inference time to mitigate the modality gap is to use the nearest text embedding as the prefix, instead of the audio embedding. We calculate the cosine similarity between the audio embedding $\mathbf{z_a}$ and the text embeddings in $\mathcal{M}$ and decode with the most similar:
\begin{equation}
    \mathbf{p} = \mathbf{z_i}  \;\; \vert \;\; i = \underset{\mathbf{z_t} \in \mathcal{M}}{argmax} \; sim(\mathbf{z_a}, \mathbf{z_t})   
\end{equation}
Where $sim(\mathbf{x},\mathbf{y})= \frac{\mathbf{x} \cdot \mathbf{y}}{\Vert \mathbf{x} \Vert  \cdot  \Vert \mathbf{y} \Vert}$. Since the decoder is trained to reconstruct the original text conditioned on the text embedding, nearest-neighbor decoding can be successful if a sufficiently similar text embedding is present in $\mathcal{M}$.

\subsubsection{Projection-based decoding}
\label{project}
A better approach is to project the audio embedding into the text embedding space. This involves obtaining the representation of the audio embedding, by combining the embeddings in $\mathcal{M}$ through a weighted combination. 
\begin{equation}
    \mathbf{p} = \sum_{i=1}^{\vert \mathcal{M} \vert} w_i * \mathbf{z_{t_i}}
\end{equation}
The weights $w_i$ for these text embeddings are determined by calculating the cosine similarity between the audio embedding $\mathbf{z_a}$ and each embedding in $\mathcal{M}$. Following \cite{decap} the similarity is then scaled by a temperature parameter $\tau$ and normalized using a softmax function:
\begin{equation}
    w_i = \frac{\exp( sim(\mathbf{z_a}, \mathbf{z_{t_i}})  / \tau)}{\sum_{j=1}^{\vert \mathcal{M} \vert} \exp( sim(\mathbf{z_a}, \mathbf{z_{t_j}}) / \tau)}
\end{equation}

\section{Experimens}
\subsection{Data}
 We conduct experiments using two benchmarks, AudioCaps and Clotho. AudioCaps contains  $~50$k, 10-second 
audio clips sourced from Audioset \cite{audioset}. Each audio is annotated
with one caption in the training set and five captions in the evaluation set. Clotho consists of 4981 audio samples of 15 to 30 seconds
duration. Each audio is annotated with five captions. We follow
the standard recipes  of training, validation, and test splits on each
dataset for our experiments. To adhere to a weakly-supervised setting we assume no access to audio data in the training and validation sets.

\subsection{Experimental setup}
To extract audio and text embeddings we employ a frozen CLAP model\footnote{https://github.com/XinhaoMei/WavCaps/tree/master} trained on WavCaps \cite{clap3}.
The audio encoder is a CNN14 from Pre-trained Audio Neural
Networks (PANNs) \cite{panns}, and the text encoder is a BERT-based model \cite{bert}.
We choose this model as the embedding extractor because AudioCaps and Clotho datasets were not included in its training set. This choice is made under the assumption that target audio data are unavailable for training purposes. The decoder $D$ consists of a mapping network $f$ which is a 2-layered MLP, and the language model which is a 4-layer Transformer \cite{transformer} with 4 attention heads. The size
of the hidden state is 768. The decoder $D$ is trained from scratch on the target captions. 
The noise variance for \textit{Noise Injection} training is set to $\sigma^2 = 0.013.$
We train the proposed model for 30 epochs using Adam optimizer \cite{adamw} and a batch size of 64. The learning rate is linearly increased
to $2 \times 10^{-5}$
in the first five epochs using warm-up, which is then
multiplied by 0.2 every 10 epochs. We use greedy search for decoding.

\begin{table*}[h]
\caption{\textbf{Results on AudioCaps and Clotho}. We report results for fully supervised methods trained on audio-caption pairs, and our proposed methods trained only on captions. \texttt{WSAC} is our baseline approach presented in Section 2. We refer to \textit{Noise injection} as \texttt{NI}, \textit{Embedding shift} as \texttt{ES}, \textit{Nearest-neighborhood decoding} as \texttt{NND} and, \textit{Projection-based decoding} as \texttt{PD}. We highlight the best results for fully and weakly supervised methods with \underline{underline} and \textbf{bold} respectively.} 
\centering
\label{tab:results}
\scalebox{0.78}{
\begin{tabular}{|cll|c|c|c|c|c|c|c|c|c|c|c|}
\hline
\multicolumn{3}{|c|}{\textbf{Dataset}} &
  \textbf{Supervision} &
  \textbf{Method} &
  \textbf{BLEU\textsubscript{1}} &
  \textbf{BLEU\textsubscript{2}} &
  \textbf{BLEU\textsubscript{3}} &
  \textbf{BLEU\textsubscript{4}} &
  \textbf{METEOR} &
  \textbf{ROUGE\textsubscript{L}} &
  \textbf{CIDEr} &
  \textbf{SPICE} &
  \textbf{SPIDEr} \\ \hline
\multicolumn{3}{|c|}{\multirow{9}{*}{Audiocaps}} &
  \multirow{4}{*}{\begin{tabular}[c]{@{}c@{}}Audio-Caption\\Pairs\end{tabular}} &
  \texttt{Mei et al.} \cite{act} &
  $0.647$ &
  $0.488$ &
  $0.356$ &
  $0.252$ &
  $0.222$ &
  $0.468$ &
  $0.679$ &
  $0.160$ &
  $0.420$ \\
\multicolumn{3}{|c|}{} &
   &
  \texttt{Kim et al.} \cite{prefix} &
  $0.713$ &
  $0.552$ &
  $0.421$ &
  $0.309$ &
  $0.240$ &
  $0.503$ &
  $0.733$&
  $0.177$ &
  $0.455$ \\
\multicolumn{3}{|c|}{} &
   &
  \texttt{Gontier et al.} \cite{bart} &
  $0.699$ &
  $0.523$ &
  $0.380$ &
  $0.266$ &
  $0.241$ &
  $0.493$ &
  $0.753$ &
  $0.176$ &
  $0.465$ \\
\multicolumn{3}{|c|}{} &
   &
  \texttt{Mei et al.} \cite{clap3} &
  $\underline{0.707}$ &
  - &
  - &
  $\underline{0.283}$ &
  $\underline{0.250}$ &
  $\underline{0.507}$ &
  $\underline{0.787}$ &
  $\underline{0.182}$ &
  $\underline{0.485}$ \\ \cline{4-14} 
\multicolumn{3}{|c|}{} &
  \multirow{5}{*}{\begin{tabular}[c]{@{}c@{}}Captions\\ Only\end{tabular}} &
  \texttt{WSAC} &
  $0.574$ &
  $0.398$ &
  $0.267$ &
  $0.167$ &
  $0.222$ &
  $0.426$ &
  $0.493$ &
  $0.155$ &
  $0.324$ \\
\multicolumn{3}{|c|}{} &
   &
  \texttt{WSAC+NI} &
  $0.662$&
  $0.477$ &
  $0.328$ &
  $0.216$ &
  $0.223$ &
  $0.46$ &
  $0.579$ &
  $0.155$ &
  $0.367$ \\
\multicolumn{3}{|c|}{} &
   &
  \texttt{WSAC+ES} &
  $0.653$ &
  $0.458$ &
  $0.300$ &
  $0.185$ &
  $0.214$ &
  $0.451$ &
  $0.540$ &
  $0.154$ &
  $0.347$ \\
\multicolumn{3}{|c|}{} &
   &
  \texttt{WSAC+NND} &
  $0.643$ &
  $0.457$ &
  $0.312$ &
  $0.198$ &
  $0.231$ &
  $0.454$ &
  $0.548$ &
  $0.166$ &
  $0.357$ \\
\multicolumn{3}{|c|}{} &
   &
  \texttt{WSAC+PD} &
  $\mathbf{0.698}$ &
  $\mathbf{0.511}$ &
  $\mathbf{0.357}$ &
  $\mathbf{0.232}$ &
  $\mathbf{0.241}$ &
  $\mathbf{0.479}$ &
  $\mathbf{0.633}$ &
  $\mathbf{0.173}$ &
  $\mathbf{0.403}$ \\ \hline
\multicolumn{3}{|c|}{\multirow{9}{*}{Clotho}} &
  \multirow{4}{*}{\begin{tabular}[c]{@{}c@{}}Audio-Caption\\Pairs\end{tabular}} &
  \texttt{Xu et al.} \cite{xu}&
  $0.556$ &
  $0.363$ &
  $0.242$ &
  $0.159$&
  $0.169$ &
  $0.368$ &
  $0.377$ &
  $0.115$ &
  $0.246$ \\
\multicolumn{3}{|c|}{} &
   &
  \texttt{Koh et al.} \cite{koh} &
  $0.551$ &
  $0.369$ &
  $0.252$ &
  $0.168$ &
  $0.165$ &
  $0.373$ &
  $0.380$ &
  $0.111$ &
  $0.246$ \\
\multicolumn{3}{|c|}{} &
   &
  \texttt{Kim et al.} \cite{prefix} &
  $0.560$ &
  $0.376$ &
  $0.253$ &
  $0.160$ &
  $0.170$ &
  $0.378$ &
  $0.392$ &
  $0.118$ &
  $0.255$ \\
\multicolumn{3}{|c|}{} &
   &
  \texttt{Mei et al.} \cite{clap3} &
  $\underline{0.601}$ &
  - &
  - &
  $\underline{0.180}$ &
  $\underline{0.185}$ &
  $\underline{0.400}$ &
  $\underline{0.488}$ &
  $\underline{0.133}$ &
  $\underline{0.310}$ \\ \cline{4-14} 
\multicolumn{3}{|c|}{} &
  \multirow{5}{*}{\begin{tabular}[c]{@{}c@{}}Captions\\Only\end{tabular}} &
  \texttt{WSAC} &
  $0.462$ &
  $0.282$ &
  $0.173$ &
  $0.102$ &
  $0.166$ &
  $0.343$ &
  $0.265$ &
  $0.113$ &
  $0.189$ \\
\multicolumn{3}{|c|}{} &
   &
  \texttt{WSAC+NI} &
  $0.525$ &
  $0.314$ &
  $0.193$ &
  $0.118$ &
  $0.164$ &
  $0.352$ &
  $0.315$ &
  $0.113$ &
  $0.214$ \\
\multicolumn{3}{|c|}{} &
   &
  \texttt{WSAC+ES} &
  $\mathbf{0.546}$ &
  $\mathbf{0.332}$ &
  $\mathbf{0.203}$ &
  $\mathbf{0.120}$ &
  $0.159$ &
  $0.353$ &
  $0.301$ &
  $0.109$ &
  $0.205$ \\
\multicolumn{3}{|c|}{} &
   &
  \texttt{WSAC+NND} &
  $0.498$ &
  $0.294$ &
  $0.179$ &
  $0.106$ &
  $0.166$ &
  $0.338$ &
  $0.332$ &
  $0.113$ &
  $0.222$ \\
\multicolumn{3}{|c|}{} &
   &
  \texttt{WSAC+PD} &
  $0.532$ &
  $0.324$ &
  $0.200$ &
  $0.118$ &
  $\mathbf{0.174}$ &
  $\mathbf{0.354}$ &
  $\mathbf{0.371}$ &
  $\mathbf{0.123}$ &
  $\mathbf{0.247}$ \\ \hline
\end{tabular}
}
\vspace{-10pt}
\end{table*}

\subsection{Compared methods and evaluation metrics}
Since no previous work has addressed AAC in similar supervision settings we compare our methods against fully supervised approaches trained on paired data. \texttt{Koh et al.} \cite{koh} use  a  latent space similarity objective and train a model with a PANNs encoder and a transformer decoder.  \texttt{Xu et al.} \cite{xu} design a GRU for the decoder. \texttt{Mei et al.} \cite{act} propose a full transformer encoder-decoder
architecture.  \texttt{Gontier et al.}  \cite{bart} utilize a pre-trained
language model based on BART \cite{bart}, and finetune it for AAC using guidance from Audioset tags. \texttt{Kim et al.} \cite{prefix} propose prefix tuning for AAC learning a prefix to guide the caption generation of a frozen GPT-2 \cite{gpt2}. \texttt{Mei et al.} \cite{clap3} utilize a CLAP audio encoder pre-trained on WavCaps and a BART decoder achieving state-of-the-art results in both Clotho and AudioCaps.
All the methods in this work are evaluated by the metrics widely
used in the captioning tasks, including BLEU \cite{bleu}, METEOR \cite{meteor},
ROUGE-L \cite{rouge}, CIDEr \cite{cider}, SPICE \cite{spice}, and SPIDEr \cite{spider}.

\subsection{Results and Discussion}
In this section, we present the results of our proposed methods on the performance metrics and compare them with fully supervised arts. 
Additionally, we illustrate the effectiveness of each strategy in reducing the modality gap. As shown in Table \ref{tab:results} our methods demonstrate comparable performance to prior state-of-the-art models despite never encountering in-domain audio data during training. We present the results of our baseline approach described in Section 2 and the results of the baseline approach in conjunction with the strategies presented in Section 3. It is evident that all the strategies boost the performance of our baseline approach in both evaluation sets. Interestingly the \textit{inference strategies} outperform the \textit{training strategies} in most cases. We hypothesize that this is because they utilize the \textit{Memory} $\mathcal{M}$ which consists of in-domain text embeddings in order to bridge the modality gap. Our best-performing method, namely \textit{Projection-based decoding} achieves 80\% and 83\% of the SPIDEr performance of the current  fully supervised state-of-the model in Clotho and AudioCaps evaluation sets respectively. Additionally \textit{Projection-based decoding} matches the performance of the of fully-supervised approaches proposed by Kim et al. \cite{prefix}. Koh et al. \cite{koh} and Xu et al. \cite{xu} in the Clotho evaluation set.

\noindent \textbf{Visualization of embeddings:} To further examine the effectiveness of the proposed strategies we illustrate the embeddings  in 2D space using t-SNE in Fig. 3. In Fig. 3a and 3b we randomly sample audio and text embeddings from the Clotho training set after applying \textit{Noise Injection} and \textit{Embedding Shift} to the text embeddings. Fig. 3c and 3d illustrate randomly selected text embeddings from the Clotho evaluation set, alongside the embeddings utilized for decoding, namely the nearest neighbors and the projections, rather than the paired audio embeddings.
It is evident that all strategies are effective in condensing the modality gap showcased in Fig. 2, where the audio and text modalities are embedded at arm’s length in their shared representation space.

\begin{figure}[t]
        \centering
        \label{vis}

        \begin{subfigure}[b]{0.22\textwidth}
            \centering
            \includegraphics[width=\textwidth]{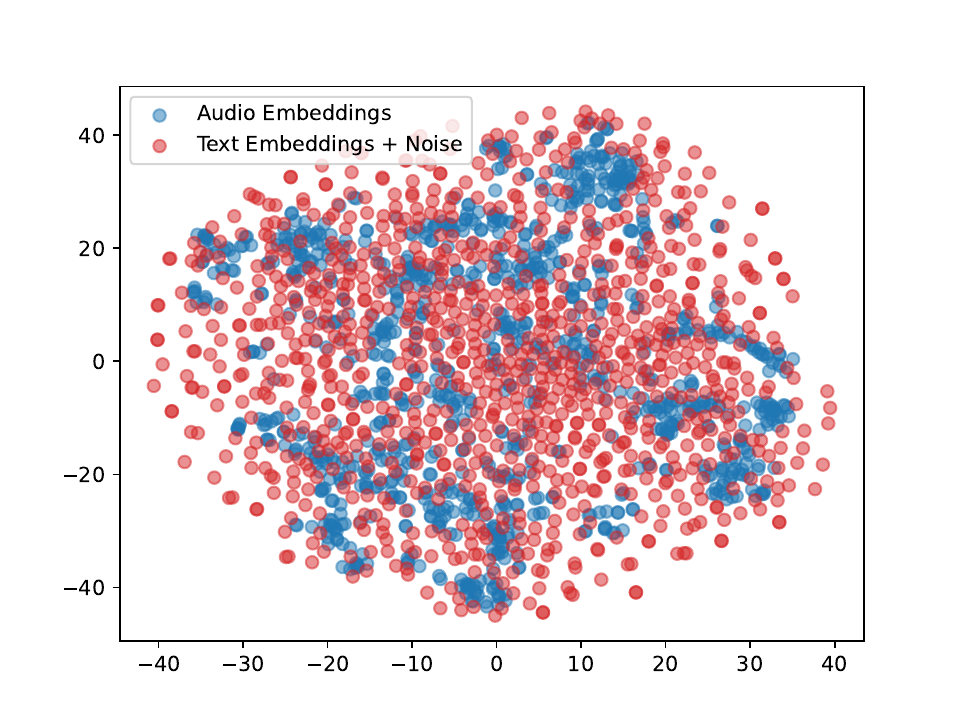}
            \caption[Noise Injection]%
            {{\small Noise Injection}}    
            \label{fig:noise}
            \vspace{-10pt}
        \end{subfigure}
        \hfill
        \begin{subfigure}[b]{0.22\textwidth}  
            \centering 
            \includegraphics[width=\textwidth]{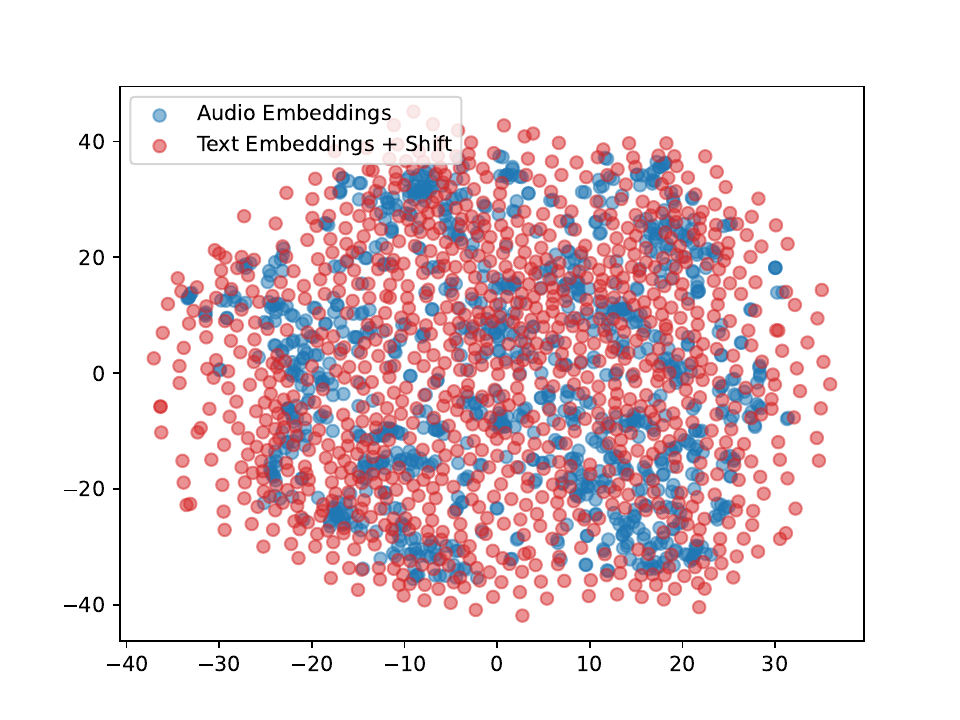}
            \caption[Embedding Shift]%
            {{\small Embedding Shift}}    
            \label{fig:shift}
            \vspace{-10pt}
        \end{subfigure}
        \vskip\baselineskip
        \begin{subfigure}[b]{0.22\textwidth}   
            \centering 
            \includegraphics[width=\textwidth]{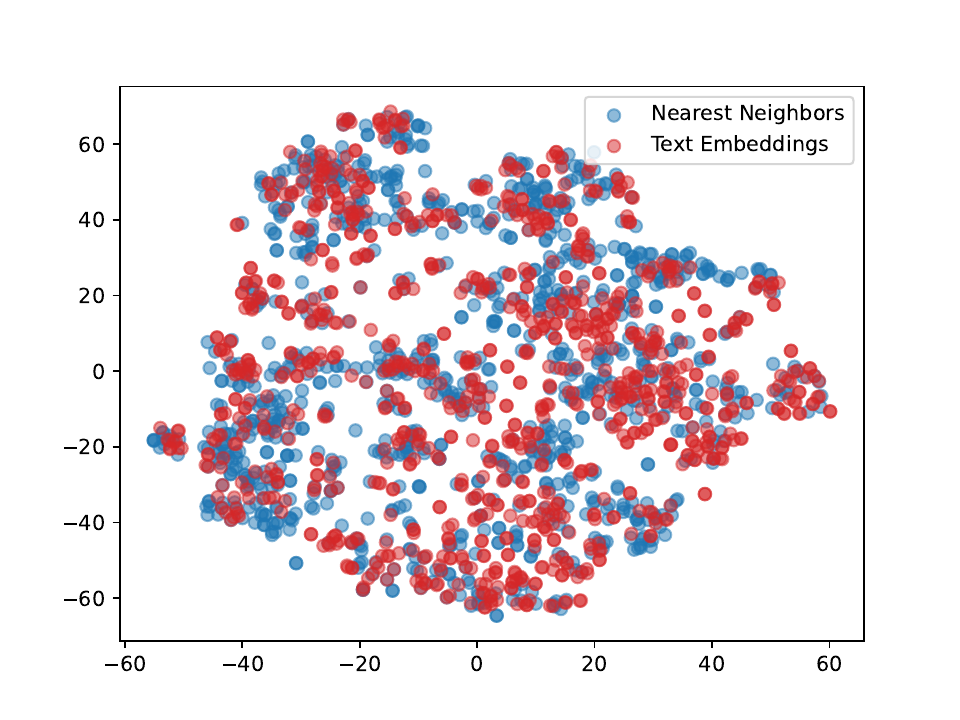}
            \caption[Nearest-neighbor decoding]%
            {{\small Nearest-neighbor decoding}}    
            \label{fig:nnd}
        \end{subfigure}
        \hfill
        \begin{subfigure}[b]{0.22\textwidth}   
            \centering 
            \includegraphics[width=\textwidth]{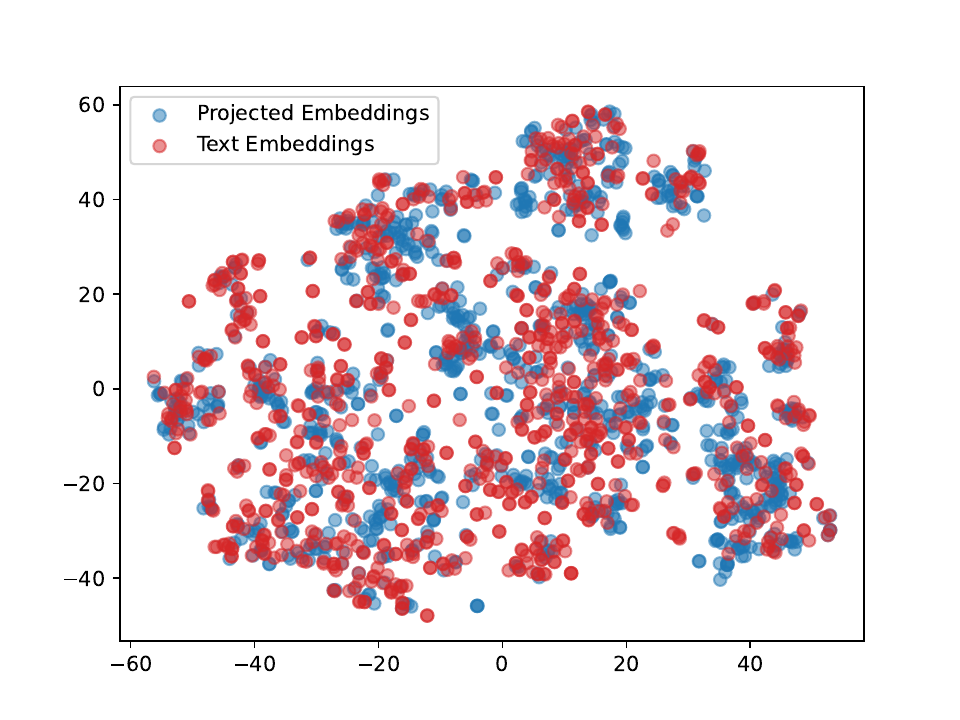}
            \caption[Projection-based decoding]%
            {{\small Projection-based decoding}}    
            \label{fig:pr}
        \end{subfigure}

        \label{fig:mean and std of nets}
        \caption[TSN-E visualizations of the embedding space after applying the strategies presented in Section 3. ]
        {\small TSN-E visualizations of the embedding space after applying the strategies presented in Section 3.} 
        
        \vspace{-15pt}
    \end{figure}

\section{Conclusion and Feature Work}
\label{sec:ack}

In this work, we propose a weakly-supervised approach for Automated Audio Captioning that requires a pre-trained CLAP model and only additional text data to train on a target domain. 
Our method alleviates the necessity of paired data in a target domain, which are hard to collect. We demonstrate that by leveraging the shared embedding space of CLAP 
we can learn to reconstruct the text from the CLAP text embedding and during inference decode using the audio embeddings. We show that such an approach is suboptimal due to the presence of a modality gap and adopt strategies that effectively mitigate it. Our best-performing method achieves comparable results to prior arts trained in a fully supervised manner. For future work, we plan to study the effectiveness of our proposed approach on other tasks, such as Music Captioning and Audio Question Answering. We further aim to train a mapping network to learn the gap between the two modalities in a supervised manner. 
\bibliographystyle{IEEEtran}
\bibliography{template}

\begin{thebibliography}{10}
\providecommand{\url}[1]{#1}
\def\UrlFont{\rmfamily}
\providecommand{\newblock}{\relax}
\providecommand{\bibinfo}[2]{#2}
\providecommand\BIBentrySTDinterwordspacing{\spaceskip=0pt\relax}
\providecommand\BIBentryALTinterwordstretchfactor{4}
\providecommand\BIBentryALTinterwordspacing{\spaceskip=\fontdimen2\font plus
\BIBentryALTinterwordstretchfactor\fontdimen3\font minus
  \fontdimen4\font\relax}
\providecommand\BIBforeignlanguage[2]{{%
\expandafter\ifx\csname l@#1\endcsname\relax
\typeout{** WARNING: IEEEtran.bst: No hyphenation pattern has been}%
\typeout{** loaded for the language `#1'. Using the pattern for}%
\typeout{** the default language instead.}%
\else
\language=\csname l@#1\endcsname
\fi
#2}}

\bibitem{clotho}
K.~Drossos, S.~Lipping, and T.~Virtanen, ``Clotho: an audio captioning
  dataset,'' \emph{ICASSP 2020 - 2020 IEEE International Conference on
  Acoustics, Speech and Signal Processing (ICASSP)}, pp. 736--740, 2019.

\bibitem{audiocaps}
\BIBentryALTinterwordspacing
C.~D. Kim, B.~Kim, H.~Lee, and G.~Kim, ``{A}udio{C}aps: Generating captions for
  audios in the wild,'' in \emph{In Proc. NAACL}.\hskip 1em plus 0.5em minus
  0.4em\relax Minneapolis, Minnesota: Association for Computational
  Linguistics, June 2019, pp. 119--132. [Online]. Available:
  \url{https://aclanthology.org/N19-1011}
\BIBentrySTDinterwordspacing

\bibitem{coala}
X.~Favory, K.~Drossos, T.~Virtanen, and X.~Serra, ``Coala: Co-aligned
  autoencoders for learning semantically enriched audio representations,''
  \emph{arXiv preprint arXiv:2006.08386}, 2020.

\bibitem{audioclip}
A.~Guzhov, F.~Raue, J.~Hees, and A.~R. Dengel, ``Audioclip: Extending clip to
  image, text and audio,'' \emph{ICASSP 2022 - 2022 IEEE International
  Conference on Acoustics, Speech and Signal Processing (ICASSP)}, pp.
  976--980, 2021.

\bibitem{clap1}
B.~Elizalde, S.~Deshmukh, M.~A. Ismail, and H.~Wang, ``Clap learning audio
  concepts from natural language supervision,'' in \emph{ICASSP 2023 - 2023
  IEEE International Conference on Acoustics, Speech and Signal Processing
  (ICASSP)}, 2023, pp. 1--5.

\bibitem{clap2}
Y.~Wu, K.~Chen, T.~Zhang, Y.~Hui, T.~Berg-Kirkpatrick, and S.~Dubnov,
  ``Large-scale contrastive language-audio pretraining with feature fusion and
  keyword-to-caption augmentation,'' \emph{In Proc. ICASSP}, vol.
  abs/2211.06687, 2022.

\bibitem{clap3}
X.~Mei, C.~Meng, H.~Liu, Q.~Kong, T.~Ko, C.~Zhao, M.~. Plumbley, Y.~Zou, and
  W.~Wang, ``Wavcaps: A chatgpt-assisted weakly-labelled audio captioning
  dataset for audio-language multimodal research,'' \emph{ArXiv}, vol.
  abs/2303.17395, 2023.

\bibitem{coco}
\BIBentryALTinterwordspacing
X.~Chen, H.~Fang, T.~Lin, R.~Vedantam, S.~Gupta, P.~Doll{\'{a}}r, and C.~L.
  Zitnick, ``Microsoft {COCO} captions: Data collection and evaluation
  server,'' \emph{CoRR}, vol. abs/1504.00325, 2015. [Online]. Available:
  \url{http://arxiv.org/abs/1504.00325}
\BIBentrySTDinterwordspacing

\bibitem{prefix}
M.-K. Kim, K.~Sung‐Bin, and T.-H. Oh, ``Prefix tuning for automated audio
  captioning,'' \emph{In Proc. ICASSP 2023}, vol. abs/2303.17489, 2023.

\bibitem{noise}
D.~Nukrai, R.~Mokady, and A.~Globerson, ``Text-only training for image
  captioning using noise-injected clip,'' in \emph{Conference on Empirical
  Methods in Natural Language Processing}, 2022.

\bibitem{decap}
W.~Li, L.~Zhu, L.~Wen, and Y.~Yang, ``Decap: Decoding clip latents for
  zero-shot captioning via text-only training,'' \emph{In Proc. ICLR}, vol.
  abs/2303.03032, 2023.

\bibitem{clip}
A.~Radford, J.~W. Kim, C.~Hallacy, A.~Ramesh, G.~Goh, S.~Agarwal, G.~Sastry,
  A.~Askell, P.~Mishkin, J.~Clark, G.~Krueger, and I.~Sutskever, ``Learning
  transferable visual models from natural language supervision,'' in \emph{In
  Proc ICML}, 2021.

\bibitem{gap}
W.~Liang, Y.~Zhang, Y.~Kwon, S.~Yeung, and J.~Y. Zou, ``Mind the gap:
  Understanding the modality gap in multi-modal contrastive representation
  learning,'' \emph{ArXiv}, vol. abs/2203.02053, 2022.

\bibitem{tsne}
\BIBentryALTinterwordspacing
L.~van~der Maaten and G.~Hinton, ``Visualizing data using {t-SNE},''
  \emph{Journal of Machine Learning Research}, vol.~9, pp. 2579--2605, 2008.
  [Online]. Available: \url{http://www.jmlr.org/papers/v9/vandermaaten08a.html}
\BIBentrySTDinterwordspacing

\bibitem{audioset}
J.~F. Gemmeke, D.~P.~W. Ellis, D.~Freedman, A.~Jansen, W.~Lawrence, R.~C.
  Moore, M.~Plakal, and M.~Ritter, ``Audio set: An ontology and human-labeled
  dataset for audio events,'' in \emph{2017 IEEE International Conference on
  Acoustics, Speech and Signal Processing (ICASSP)}, 2017, pp. 776--780.

\bibitem{panns}
Q.~Kong, Y.~Cao, T.~Iqbal, Y.~Wang, W.~Wang, and M.~D. Plumbley, ``Panns:
  Large-scale pretrained audio neural networks for audio pattern recognition,''
  \emph{IEEE/ACM Transactions on Audio, Speech, and Language Processing},
  vol.~28, pp. 2880--2894, 2019.

\bibitem{bert}
J.~Devlin, M.-W. Chang, K.~Lee, and K.~Toutanova, ``{BERT}: Pre-training of
  deep bidirectional transformers for language understanding,'' in \emph{In
  Proc. ACL, Volume 1 (Long and Short Papers)}, June 2019.

\bibitem{transformer}
A.~Vaswani, N.~M. Shazeer, N.~Parmar, J.~Uszkoreit, L.~Jones, A.~N. Gomez,
  L.~Kaiser, and I.~Polosukhin, ``Attention is all you need,'' in \emph{NIPS},
  2017.

\bibitem{adamw}
I.~Loshchilov and F.~Hutter, ``Fixing weight decay regularization in adam,''
  \emph{ArXiv}, vol. abs/1711.05101, 2017.

\bibitem{act}
X.~Mei, X.~Liu, Q.~Huang, M.~D. Plumbley, and W.~Wang, ``Audio captioning
  transformer,'' in \emph{DCASE Workshop}, 2021.

\bibitem{bart}
F.~Gontier, R.~Serizel, and C.~Cerisara, ``Automated audio captioning by
  fine-tuning bart with audioset tags,'' in \emph{DCASE Workshop}, 2021.

\bibitem{xu}
X.~Xu, H.~Dinkel, M.~Wu, Z.~Xie, and K.~Yu, ``Investigating local and global
  information for automated audio captioning with transfer learning,'' \emph{In
  Proc. ICASSP}, pp. 905--909, 2021.

\bibitem{koh}
A.~Koh, X.~Fuzhao, and C.~E. Siong, ``Automated audio captioning using transfer
  learning and reconstruction latent space similarity regularization,'' in
  \emph{In Proc. ICASSP}.\hskip 1em plus 0.5em minus 0.4em\relax IEEE, 2022,
  pp. 7722--7726.

\bibitem{gpt2}
A.~Radford, J.~Wu, R.~Child, D.~Luan, D.~Amodei, and I.~Sutskever, ``Language
  models are unsupervised multitask learners,'' 2019.

\bibitem{bleu}
\BIBentryALTinterwordspacing
K.~Papineni, S.~Roukos, T.~Ward, and W.-J. Zhu, ``{B}leu: a method for
  automatic evaluation of machine translation,'' in \emph{ACL}.\hskip 1em plus
  0.5em minus 0.4em\relax Philadelphia, Pennsylvania, USA: Association for
  Computational Linguistics, July 2002, pp. 311--318. [Online]. Available:
  \url{https://aclanthology.org/P02-1040}
\BIBentrySTDinterwordspacing

\bibitem{meteor}
\BIBentryALTinterwordspacing
S.~Banerjee and A.~Lavie, ``{METEOR}: An automatic metric for {MT} evaluation
  with improved correlation with human judgments,'' in \emph{ACL}.\hskip 1em
  plus 0.5em minus 0.4em\relax Ann Arbor, Michigan: Association for
  Computational Linguistics, June 2005, pp. 65--72. [Online]. Available:
  \url{https://aclanthology.org/W05-0909}
\BIBentrySTDinterwordspacing

\bibitem{rouge}
\BIBentryALTinterwordspacing
C.-Y. Lin, ``{ROUGE}: A package for automatic evaluation of summaries,'' in
  \emph{Text Summarization Branches Out}.\hskip 1em plus 0.5em minus
  0.4em\relax Barcelona, Spain: Association for Computational Linguistics, July
  2004, pp. 74--81. [Online]. Available:
  \url{https://aclanthology.org/W04-1013}
\BIBentrySTDinterwordspacing

\bibitem{cider}
R.~Vedantam, C.~L. Zitnick, and D.~Parikh, ``Cider: Consensus-based image
  description evaluation,'' in \emph{In Proc. CVPR}, 2015, pp. 4566--4575.

\bibitem{spice}
P.~Anderson, B.~Fernando, M.~Johnson, and S.~Gould, ``Spice: Semantic
  propositional image caption evaluation,'' in \emph{In Proc ECCV}.\hskip 1em
  plus 0.5em minus 0.4em\relax Springer, 2016, pp. 382--398.

\bibitem{spider}
S.~Liu, Z.~Zhu, N.~Ye, S.~Guadarrama, and K.~P. Murphy, ``Optimization of image
  description metrics using policy gradient methods,'' \emph{ArXiv}, vol.
  abs/1612.00370, 2016.

\end{thebibliography}

%
%
%
%
%
%
%
%
%

\end{sloppy}
\end{document}